\documentclass{egpubl}
\usepackage{egsgp2026}
\usepackage{algorithm}
\usepackage{algpseudocode}
 
\SpecialIssueSubmission %

\usepackage[T1]{fontenc}
\usepackage{dfadobe}  

\usepackage{cite}  %
\BibtexOrBiblatex
\electronicVersion
\PrintedOrElectronic
\ifpdf \usepackage[pdftex]{graphicx} \pdfcompresslevel=9
\else \usepackage[dvips]{graphicx} \fi

\usepackage{egweblnk} 

\title[Progressive Convex Hull Simplification]{Progressive Convex Hull Simplification\vspace*{-0.2in}}

\author[\censor{A. Jacobson}]
{\parbox{\textwidth}{\centering \censor{Alec Jacobson}}
        \\
{\parbox{\textwidth}{%
\centering%
\censor{University of Toronto}%
       }
}
}

\usepackage{amsmath}
\usepackage{amssymb}
\usepackage{wrapfig}
\usepackage{xcolor}

\newcommand{\newhl}[1]{#1}
\newcommand{\censor}[1]{#1}

\usepackage{xspace}
\newcommand{\Vrep}{V-representation\xspace}
\newcommand{\Hrep}{H-representation\xspace}
\newcommand{\x}{\mathbf{x}}
\newcommand{\n}{\mathbf{n}}
\renewcommand{\c}{\mathbf{c}}
\renewcommand{\v}{\mathbf{v}}

\begin{document}

\teaser{
 \vspace*{-0.2in}
 \includegraphics[width=0.9\linewidth]{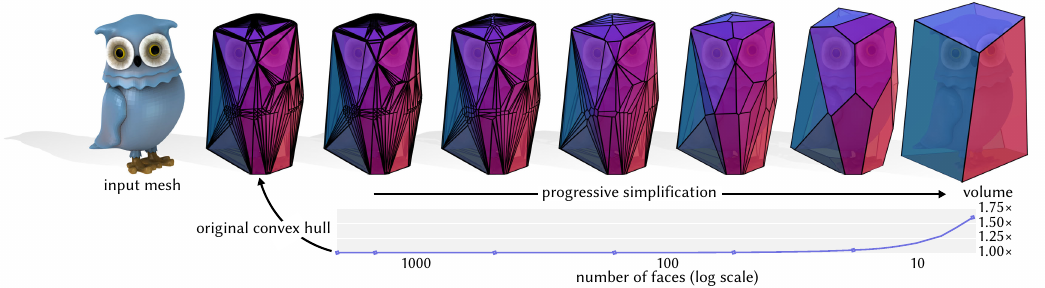}
 \centering
  \caption{
    Our convex hull simplification iteratively removes the next halfspace that would add the least amount of volume.
    This results in very tight fitting hulls (under 5\% increase) down to a small number of faces (18). Even for an extreme simplification (6 faces), the simplified hull is a modestly tighter fit than an oriented bounding box (1.61× vs. 1.72×).
  }
\label{fig:owl}
}

\maketitle

\begin{abstract}
Convex hulls are useful as tight bounding proxies for a variety of tasks including collision detection, ray intersection, and distance computation. Unfortunately, the complexity of polyhedral convex hulls grows linearly with their input. We consider the problem of conservatively simplifying a convex hull to a specified number of half-spaces while minimizing added volume or surface area. By working in the dual representation, we propose an efficient $O(n \log n)$ greedy optimization. In comparisons, we show that existing methods either exhibit poor efficiency, tightness or safety. We demonstrate the success of our method on a variety of input shapes and downstream application domains.
\end{abstract}

\section{Introduction}
\begin{figure}[b!]
\vspace*{-0.2in}
    \includegraphics[width=\linewidth]{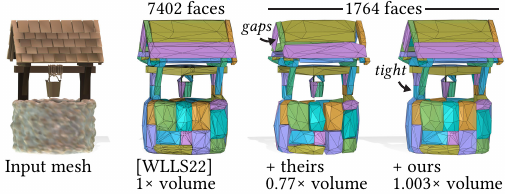}
    \caption{%
    Our simplification is a natural drop-in to approximate convex decomposition methods, e.g., CoACD \cite{wei2022coacd}.
    Their software optionally allows decimation (vertex dropping) as a post-process, but results in gaps between hulls, which permit false negatives during collision handling. Our simplification is tight and safe.
    }
    \label{fig:coacd}
\end{figure}

Convex hulls are --- by definition --- the tightest convex bounding volume of a shape.
This makes them useful as primitives in bounding volume hierarchies or directly as broadphase proxies for a variety of shape query tasks including ray intersection, collision detection \& handling, distance computation, and sampling.
Unfortunately, the complexity of a polyhedral convex hull grows linearly with that of the shape it encloses: a surface mesh with $n$ triangles 
will generally induce a convex hull with $O(n)$ faces. For example, the convex hull of the modest triangle mesh in Fig.~\ref{fig:owl} has over 2,000 faces.
Compared to the fixed-complexity of simpler bounding boxes or $k$-DOPs, the complexity of convex hulls slows down their promise of acceleration as a bounding volume.
Convex hull simplification methods exist, but their simplified hulls are either no longer guaranteed to contain the input shape or 
sacrifice the tightness of the hull by adding too much volume or surface area.
These issues are compounded for convex hulls of subparts of a shape during approximate convex decomposition (see Fig.~\ref{fig:coacd}).

In this paper, we build
on the theoretical results promising the effectiveness of successive halfspace removal (e.g., \cite{Reisner2001,LopezR00,LopezR02}) and 
present a practical conservative convex hull simplification heuristic
based on greedy halfspace removal using a priority queue in the style of
triangle mesh simplification
\cite{Hoppe96,GarlandH97}.
By utilizing the duality of convex polyhedra, the cost of each
candidate halfspace removal is simple to compute robustly and efficient (requiring only local information).
The simplified hull conservatively contains the input hull at every step.
Our simplified convex hulls are significantly tighter than alternative heuristics.
Our optimization lends itself to customization, e.g.,  favoring minimum surface-area instead of minimum-volume, constraining selected faces on the output, inner approximations instead of outer.
We also demonstrate the effectiveness of our simplification as part of bounding volume hierarchy and approximate convex decomposition workflows.
Our source code is public at \href{https://github.com/alecjacobson/progressive-convex-hull-simplification}{https://github.com/alecjacobson/progressive-convex-hull-simplification}.

\section{Background}
We are interested in polyhedral convex hulls bounding a finite set of points or a triangulated mesh. These convex hulls are polyhedra with finite vertices connected by (possibly non-triangular) faces.
Methods for hull simplification differ primarily based on whether the hull is represented by its supporting vertices (\Vrep) or as a set of intersecting halfspaces corresponding to each face-plane (\Hrep).
Before discussing related work, we lay out some foundations.

Common convex hull algorithms (e.g., \cite{BarberDH96,ClarksonS89}) and implementations (e.g., \textsc{CGAL}, \textsc{QHull}) work with \Vrep{s} for input and output: given as input a set of 3D points, return as output a set of faces indexing those vertices.
However,
downstream applications will often make use of the output's \Hrep, converting each face $i$ to a plane equation defined by
 $\x \cdot \n_i + b_i \le 0$, where $\n_i$ and $b_i$ are the normal to the plane and its distance (as a multiple of $\n_i$) from the origin.
For example, to test if two convex shapes overlap we can gather the halfspaces of their \Hrep{s} and determine if the total intersection is empty by solving a pure-feasibility linear program.
\begin{align}
\text{find } \x \in \mathbb{R}^3 \quad \text{subject to } \n_i \cdot \x \le -b_i \quad \forall i
\end{align}
where $i$ runs over the total $m$ halfspaces gathered from both convex shapes. This fixed dimension problem can be solved very efficiently in $O(m)$ time \cite{Seidel91}.

\begin{wrapfigure}{r}{1.0in}
\includegraphics[width=1.0in,trim=0.5cm 0 0 0.5cm]{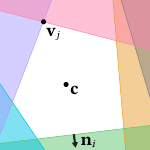}
\end{wrapfigure}
As we've seen, any \Vrep \emph{output} hull can be easily converted to a \Hrep. 
It's also straightforward to bootstrap any 
convex hull algorithm to take halfspaces as 
\emph{input} by leveraging convex duals (see, e.g., 
\cite{Gruenbaum1967}).
Suppose we're given a set of $n$ halfspaces (see 2D inset).
We can compute the convex set forming the intersection of these halfspaces by first choosing a center point $\c \in \mathbb{R}^3$ known to lie inside the resulting intersection.

For each \emph{primal halfspace}, we construct its \emph{dual vertex} about this center point via:
\begin{equation}
   \mathbf{\phi}_i = \frac{-\n_i}{\n_i \cdot \c + b_i}.
   \label{equ:todual}
\end{equation}

Now, we run any \Vrep convex hull algorithm on these 
$\phi_i$ vertices, 
producing a list of face indices referencing them (see 2D inset).
Dual vertices incident on these faces correspond 
\begin{wrapfigure}{r}{1.0in}
\includegraphics[width=1.0in,trim=0.5cm 0 0 0.5cm]{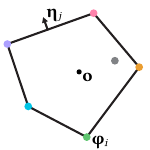}
\end{wrapfigure}
to primal halfspaces contributing to the intersection's surface; unreferenced (interior) dual vertices correspond to halfspaces too far outside the intersection to matter (grey in insets).
Each dual face's plane defines a \emph{dual halfspace}, corresponding to a \emph{primal vertex}, via:
\begin{equation}
   \v_j = \c - \frac{\mathbf{\eta}_j}{\beta_j},
   \label{equ:fromdual}
\end{equation}
where $\mathbf{\eta}_j$ and $\beta_j$ are the normal and distance (as a multiple of $\eta_j$) from the origin of the $j$th dual halfspace.
The connectivities of the primal and dual hulls are dual in the typical fashion \cite{Gruenbaum1967,GuibasS85,MeyerDSB02,JuLSW02}: i.e., edges between vertices of the primal mesh correspond to face-adjacency in the dual mesh, and \emph{vice versa}.

Exploiting the dual-representation allows construction of the polyhedral intersection of primal halfspaces without any delicate intersection constructions or mesh arrangements (e.g., \cite{Zhou:2016:MASG}).
Common convex hull implementations (e.g., \textsc{CGAL}, \textsc{QHull}) offer halfspace intersection routines implemented in this dual fashion.

The choice of center point $\c$ doesn't matter so long as it's interior. 
In practice, Eqs.~\ref{equ:todual} and~\ref{equ:fromdual} become numerically sensitive as $\c$ gets too close to any halfspace boundary.
For a safe and efficiently identified choice 
(or to determine that the halfspace intersection is empty or unbounded), we can solve a linear program to construct the center of the largest inscribed sphere, the Chebyshev center:
\begin{align}
\mathop{\text{maximize}}_{\c \in \mathbb{R}^3, r \ge 0} &\quad r \\
\text{subject to } & \quad \n_i \cdot \c + \|\n_i\| r \le -b_i \quad \forall i.
\label{equ:center}
\end{align}
Once again, the dimension of this linear program is small, so we may use very efficient $O(n)$ solvers (e.g., \cite{Seidel91}).

In general position, the convex hull of a set of points will be a polyhedron with triangular faces.
Similarly, the intersection of a set of general halfspaces produces triangle faces on the dual hull. In turn, this corresponds to primal vertices with valence 3: three planes meet at a point.
Primal faces will be polygons with any number of sides.
For our algorithm, we won't need a strong general position assumption; instead, any 
arbitrarily triangulation of non-triangle dual faces will work without affecting behavior.

\begin{figure*}
    \centering
    \includegraphics[width=\linewidth]{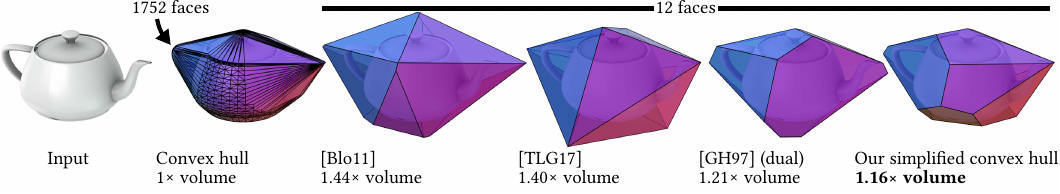}
    \caption{
    \label{fig:teapot}
    We compare the family of convex hull simplification techniques that are variants of mesh simplification \cite{GarlandH97}.
    Our method achieves the smallest volume change. Runner up is running naive appearance preserving mesh simplification on the dual hull \cite{GarlandH97} (dual).
    }
    \centering
    \includegraphics[width=\linewidth]{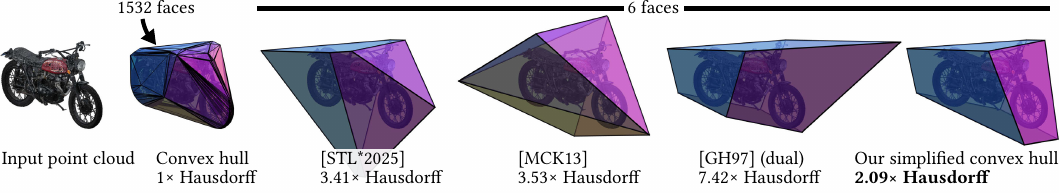}
    \caption{
    \label{fig:motorcycle}
    We take as input a point cloud, build its exact convex hull and then compare extreme simplifications with a variety of methods.
    Our method finds the tightest simplification measured not just in terms of volume and area (which we can target to minimize directly), but also in terms of Hausdorff distance measured \emph{post facto}.
    }
\end{figure*}

\section{Related Work}
Convex hull simplification has been considered in theory and practice.
A basic and key observation of many previous works is that the intersection of any subset of halfspaces in an \Hrep 
will contain the original primal hull (outer approximation).
Similarly (and useful when thinking 
about the dual hull), any convex hull of a 
subset of vertices will be contained inside the convex hull of all of the vertices (inner approximation).

\subsection{Theory and Bounds}
Lopez \& Reisner~\cite{LopezR00} began in 2D proving bounds on inner (and outer) approximations produced by greedily removing vertices (or halfspaces) according to area change.
In 2D, this area of change is always convex. In 3D and higher, dropping a vertex can produce a non-convex ``cap'' region (though its dual \emph{is} convex; see Appendix~\ref{sec:dual-caps-are-non-convex}). Overcoming this challenge, Reisner et al.~\cite{Reisner2001} prove analogous bounds for dimensions greater than two, and Lopez \& Reisner~\cite{LopezR02} attach the local bounds theoretically to a greedy algorithm.
Cast as an instance of the broader coreset selection problem, theoretical inner approximations have also been considered by forward selection approaches. \newhl{For example, for sets of points in any dimension, Klimenko~\cite{Klimenko23} considers selecting one vertex at a time and proves bounds on Hausdorff distance of the selection's convex hull to the full hull.
Similarly, Clarkson~\cite{ClarksonS89} studies reweighted random subset selections of points and bounds on distances of their hull to an input convex hull.}

\subsection{Mesh Decimation}
Convex hulls as a \Vrep are triangulatable meshes, so one approach is to just run standard edge-collapsing mesh simplification \cite{GarlandH97,Hoppe96}.
Constraining edge-collapse placement to either endpoint 
effectively drops vertices of the \Vrep,
 shrinking the hull inward.
 Vertex dropping is useful if the goal is an 
inner approximation, but for applications like collision detection this can lead to false negatives \cite{wei2022coacd} (see Fig.~\ref{fig:coacd}).
To construct an outer approximation, Bloom~\cite{bloom2011convexhull}\footnote{%
This blog post also refers to an idea attributed to Stan Melax
which sounds similar to ours but using geometric acceleration and explicit primal intersections. Unfortunately, no further details or implementation exists.} proposes to run appearance-preserving mesh simplification \cite{GarlandH97} on convex hulls and then offsetting all resulting face plane equations to contain the input. That is, simplify as a \Vrep and then ensure containment via conversion to \Hrep.
Alternatively, Tan et al.~\cite{TanLG17} collapse edges according to a locally volume non-decreasing optimization \cite{SanderGGHS00}. While their application prefers a pure \Vrep solution, it comes at the cost of containment tightness.
We show that \Vrep simplification is suitable for inner approximations, but \Hrep simplification is superior for outer approximation (see Fig.~\ref{fig:teapot}). Conveniently, primal \Hrep simplification can be mapped to dual \Vrep simplification.

As a warmup, we consider simply running appearance-preserving mesh simplification \cite{GarlandH97} on the dual mesh.
This already often outperforms previous works and enjoys the simplicity of just stringing together two common subroutines without changing them at all (convex hull, mesh simplification). But quadric error metrics in the dual space do not map
directly to volume or area minimization in the primal space.

\subsection{Forward Selection}
\begin{figure}
    \centering
    \includegraphics[width=\linewidth]{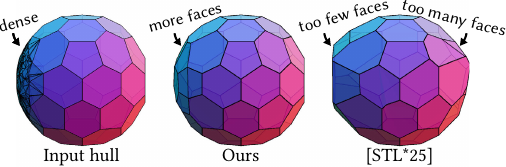}
    \caption{
    \label{fig:k-means-killer}
    A coarse hull of a sphere is augmented with a dense region, benign for our volume-minimizing simplification but exacerbates the flaw of clustering based on normals \cite{Schott25}.
    }
  \vspace*{0.2cm}
  \includegraphics[width=\linewidth]{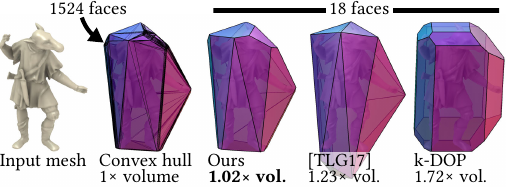}
  \caption{
  We greedily remove halfspaces from the input exact convex hull to minimize volume. Some alternatives do not limit their output to existing faces planes, but nevertheless perform inferiorly.
  }
  \label{fig:Actaeon}
  \vspace*{-0.2cm}
\end{figure}

Our proposed greedy algorithm can be treated as a backward elimination solution to the NP-hard \cite{KlimenkoRB21} subset selection problem.
An alternative approach is forward selection, inserting points into a \Vrep of halfspaces into an \Hrep, while trying to restore the original hull's volume as efficiently as possible.
In theory, Blum et al.~\cite{BlumHR19} consider iteratively adding farthest points to a running convex hull and prove bounds.
In practice, M\"uller et al.\cite{MuellerCK13} implement a similar 
algorithm, picking vertices in order of maximal recovered volume. 
Like Bloom~\cite{bloom2011convexhull}, they convert the resulting \Vrep to \Hrep for strict containment.
M\"uller et al.~do not mention acceleration; a naive implementation of their pseudocode would be $O(n k^2)$, for $O(n)$ input points and $O(k)$ targeted output faces. So, for small output acceleration may be negligible. 
Unfortunately, the resulting outputs are not particularly tight (see Fig.~\ref{fig:motorcycle}).
Schott et al.~\cite{Schott25} cluster halfspaces by normal vectors using $k$-means, refitting the \Hrep for strict containment (as \cite{bloom2011convexhull}). Unfortunately, this has the opposite of the desired effect (see Fig.~\ref{fig:k-means-killer}).

\subsection{Other Bounding Primitives}
Exact convex hulls are at the far tight-but-expensive end of the bounding volume spectrum.
On the other end, axis-aligned bounding boxes and more generally $k$-DOPs are loose fitting but make up for it with very efficient fitting and evaluation speed.
These methods are effectively \Hrep approximations of the convex hull where $m=2k$ and plane orientations ($\pm\n_i$) are fixed (typically canonical) storing just $b^+_i,b^-_i$ fit to be as tight as possible.
Finding an optimal orientation makes these representations rotation equivariant (similar to convex hulls) while introducing three degrees of freedom to further exploit for tightness.
This orientation is itself a non-linear optimization problem; \textsc{CGAL} uses a hybrid-genetic algorithm \cite{ChangGM11}, but is out-performed both in terms of tightness and speed by brute force search over well-sampled rotations \cite{Alexa22}. %
$k$-DOPs are a natural baseline to measure the success of convex hull simplification (see Fig.~\ref{fig:Actaeon}).
Other primitives, such as prisms or zonoids, have specialized or limited practical use for the tasks considered here.

\section{Method}
We take as input a set of points in $\mathbb{R}^3$, which define a convex hull.
The halfspaces defined by each face of the convex hull equivalently describe its volume as an intersection.
Our goal is to select a given target number $n$ of these halfspaces so that the resulting volume has grown as little as possible.

We immediately convert every non-degenerate face of the convex hull of the input points to its corresponding dual vertex about the hull's Chebyshev center (see Eq.~(\ref{equ:center})).
A face is considered degenerate if its area-normal vector is numerically unstable. 
Unless otherwise noted,
all subsequent steps in our algorithm are closed under rationals (e.g., \texttt{CGAL::Epeck} or \textsc{GMP}'s \texttt{mpq\_t}). In this case, using rationals means degeneracy corresponds to exactly zero area.
However, rational types are slow, so when using floating-point values, we filter out faces with tiny areas (below $10^{-8}$) and nearly identical normals (below $10^{-14}$ distance in dual space).

We initialize a triangulated convex hull of the dual vertices corresponding to each primal halfspace. We will refer to this as the ``dual hull''.
For each dual vertex, we will consider the \emph{cost} of its removal and the \emph{topological change} to 
the dual hull to maintain convexity.
Let's consider the topology change first.

\begin{wrapfigure}{r}{1.0in}
\includegraphics[width=1.0in,trim=0.5cm 0 0 0.5cm]{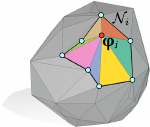}
\end{wrapfigure}
\subsection{Topology Change}
\label{sec:topology-change}
Before a dual vertex $i$ is considered for removal, we record its one-ring of incident triangles (see inset).
The boundary of this one-ring defines a topological (but geometrically not necessarily planar) polygon with vertices $\mathcal{N}_i$
Our goal is to find a triangulation of this polygon which remains outwardly convex.
We only need to consider this polygon by the convexity of the dual hull: all faces outside the one-ring are not visible to the vertex $i$ and thus their connectivity is not affected by its removal.

\begin{wrapfigure}{r}{1.0in}
\includegraphics[width=1.0in,trim=0.5cm 0 0 0.5cm]{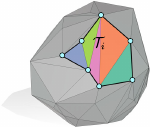}
\end{wrapfigure}
One option is to compute the convex hull of the one-ring neighbors $\mathcal{N}_i$ and only keep the outer envelope. That is, only 
keep faces $a,b,c$ such that vertex $i$ is on the positive side (according to 
$\text{\newhl{orient3D}}(\phi_a,\phi_b,\phi_c,\phi_i)$, see, \cite{shewchuk97a}).
This is a good option if $k = |\mathcal{N}_i|$ is large, as its worst case runtime is $O(k \log k)$.
Most of the time $k\approx 6$. And for those small neighborhoods, it's much faster to trivially triangulate 
$\mathcal{N}_i$ with a fan then flip non-boundary edges until all are convex (worst case $O(k^2)$).
Let's call this new local triangulation $\mathcal{T}_i$, where each $\{a,b,c\} \in \mathcal{T}_i$ has $a,b,c \in \mathcal{N}_i$.
Importantly, the boundary of $\mathcal{T}_i$ matches (in order) the one-ring $\mathcal{N}_i$.

\begin{wrapfigure}{r}{1.0in}
\includegraphics[width=1.0in,trim=0.5cm 0 0 0.5cm]{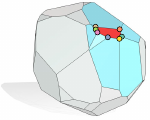}
\end{wrapfigure}
\subsection{Geometry Change}
\label{sec:geometry-change}
The cost of removing vertex $i$ is the \emph{primal} volume added by removing the corresponding primal halfspace (red face in inset) and allowing a new ``cap'' to extend as previously neighboring halfspaces intersect.
We can construct the primal geometry of this cap without explicit intersections by leveraging the dual representation and the new triangulation $\mathcal{T}_i$.
Starting with a reversed version of $\mathcal{T}_i$ we append all original triangles incident on vertex $i$.
Since the boundary of $\mathcal{T}_i$ matches $\mathcal{N}_i$, this forms a topologically closed (and genus 0) ``local dual mesh''.

\begin{wrapfigure}{r}{1.0in}
\includegraphics[width=1.0in,trim=0.5cm 0 0 0.5cm]{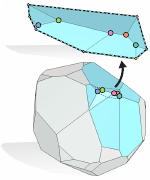}
\end{wrapfigure}
In the dual space, the corresponding mesh is not necessarily convex (see Appendix~\ref{sec:dual-caps-are-non-convex}).
Converting each face $j$ of this local dual mesh to a primal vertex (see, e.g., Eq.~(\ref{equ:todual})) and 
constructing the dual-dual (primal) connectivity, we have a local primal mesh of the ``cap'' (see inset).
This is by definition convex as it corresponds to the intersection of all primal halfspaces in $\mathcal{N}_i$ and the negated halfspace $i$.
The volume of this local primal mesh can be computed using Gauss formula (e.g., by summing the area-normal dotted with the centroid of each face).

The surface area change can similarly be computed (albeit not using rationals) by taking the primal area of faces corresponding to $\mathcal{N}_i$ minus the primal area corresponding to $i$.

\begin{wrapfigure}{r}{1.0in}
\includegraphics[width=1.0in,trim=0.5cm 0 0 0.5cm]{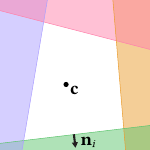}
\includegraphics[width=1.0in,trim=0.5cm 0 0 0.0cm]{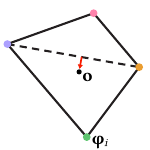}
\end{wrapfigure}
\subsection{Infinite Volume}
Sometimes removing a dual vertex $i$ corresponds to an infinite primal volume change (consider removing the green halfspace in the 2D inset).
This happens precisely when the origin in dual space falls outside the new triangulation $\mathcal{T}_i$ (next 2D inset).
This can be robustly determined by testing 
$\text{\newhl{orient3D}}(\phi_a,\phi_b,\phi_c,\mathbf{0})$ for each triangle $\{a,b,c\} \in \mathcal{T}_i$.
In practice, dual-to-primal conversion becomes numerically sensitive even when the origin is \emph{nearly} on the outside. Fortunately, this situation corresponds to either extremely large positive volume change (and will be effectively ignored during processing) or extremely negative (e.g., $-10^{20}$) volume change (which is easily identified).

\subsection{Greedy Elimination}
For each original dual vertex, we consider its removal, storing a record of its topological change $\mathcal{T}_i$ and pushing its geometric cost onto a priority queue.
When a dual vertex $i$ is popped from the queue, it is removed from the dual hull and its 
one-ring $\mathcal{N}_i$ is triangulated according to $\mathcal{T}_i$.
Maintaining the dual hull with a half-edge data structure makes this topological change efficiently output sensitive.

After dual vertex $i$ is removed, the local neighborhood of each former neighbor $j \in \mathcal{N}_i$ has changed.
Updated costs and records are computed for these dual vertices.
They're updated in the priority queue.

Dual vertices are popped from the queue until the \newhl{user's input} target number is reached.
Most applications merely require halfspace representations: each surviving dual vertex is converted to dual plane by reversing Eq.~(\ref{equ:todual}).
For visualization, or applications requiring a full primal mesh, the primal vertices are computed using Eq.~(\ref{equ:fromdual}) connected with (generally polygonal) dual connectivity.

\begin{figure}
    \includegraphics[width=\linewidth]{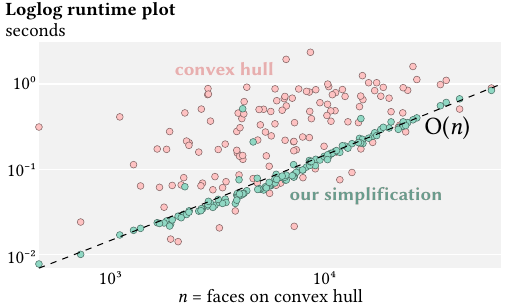}
    \caption{%
    We collect performance timings for each of the 
    133 models from threedscans.com.
    We time initial exact convex hull (pink) and our simplification to 18 half-spaces (teal).
    }
    \label{fig:threedscans-timings}
    \vspace{0.2cm}
    \includegraphics[width=\linewidth]{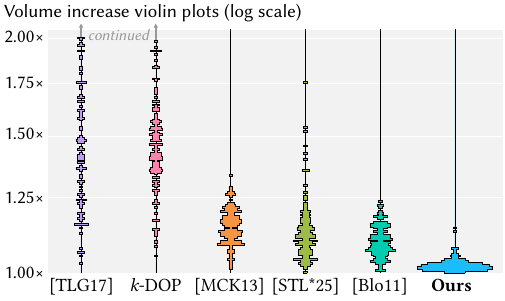}
    \caption{%
    We simplify each of the 133 models from  threedscans.com to 18 faces. We show a histogram collecting each output as a factor of the corresponding input hull volume. Our simplification was always smallest and often by a significant margin.
    }
    \label{fig:threedscans-volume-violins}
\end{figure}

\section{Implementation Details}
Our initial \textsc{Matlab} prototype used a simple dynamic adjacency matrix to track topology changes: this is completely sufficient, as strict manifoldness by construction is not needed.
In our inevitable C++ implementation, we use \textsc{CGAL}'s \texttt{Polyhedron} class to track the topology of the dual hull.
Besides being efficient, this strictly manifold half-edge data structure makes it convenient for extracting the topological dual of the dual hull (the primal hull) for final output or intermediary visualization.
Although the dual hull could have polygonal faces with more than three sides, we always triangulate. 
If there's a polygon face with consecutive vertices $ijk$ then removing $j$ only requires that $i$ and $k$ are collected. If the triangulation includes other vertices, these will be 
harmlessly retriangulated during local triangulation and any performance impact is amortized away.
While all geometric computation is compatible with rationals (e.g., \texttt{CGAL::Epeck}), we default to \texttt{CGAL::Epick} which provides exact predicate tests, specifically \texttt{orientation} (implementing \newhl{orient3D}).
We call \textsc{CGAL}'s \texttt{convex\_hull\_3} for the initial hull computations and for one-ring topology changes for high valence vertices ($\ge100$).
We implement the greedy priority queue using a lazy min heap to handle cost updates.
We use \textsc{libigl} \cite{libigl} for mesh I/O, \textsc{Polyscope} \cite{polyscope} for visualization, \textsc{SDLP} \cite{sdlp} for linear programming, and 
\textsc{Eigen} \cite{eigenweb} for linear algebra.

\section{Results \& Experiments}
We tested our implementation on a MacBook Air laptop (Apple M4 32 GB).
Like greedy mesh simplification \cite{GarlandH97}, we expect $O(n \log n)$ performance.
In Fig.~\ref{fig:threedscans-timings}, we report performance timings on all 133 models from threedscans.com.
As a baseline, we show the time to extract the initial primal hull of the input model's vertices: wide variance and as expected not very output sensitive but still roughly $O(n \log n)$ performance.
Then we time simplifying every hull down to 18 halfspaces (dual vertices). Here we see tightly linear performance matching the $O(n \log n)$ expectation.
\newhl{The scattered teal outliers tend to be inputs with extremely large valence faces (e.g., a cylinder like base with 500+ neighbors), an edge-case for our optimizations in Section~\ref{sec:topology-change}.}
The median time to simplify was 32\% of the time to compute the corresponding input hull.
All simplifications took less than a second with a  median absolute time of 81 milliseconds.

In Fig.~\ref{fig:threedscans-volume-violins}, we compare the volume gained by our simplification on this dataset against four previous works (all targeting 18 output faces) and $k$-DOPs using $18$ canonical directions as a baseline.
The violin plots show histograms of output volume relative to the input convex hull's volume on a log scale.
For every model, our output had smallest volume. 
The ``simplify-then-offset'' technique proposed by Bloom~\cite{bloom2011convexhull} performs second best, but with a median volume increase of $1.09\times$ versus our $1.02\times$.
The clustering \cite{Schott25} and forward selection \cite{MuellerCK13} perform reasonably, while \Vrep primal mesh simplification \cite{TanLG17} performs worse than $k$-DOPs.

In Fig.~\ref{fig:teapot}, we compare to previous works that are also variants of greedy mesh simplification \cite{GarlandH97}.
Simplifying the teapot's hull down to 12 faces, our method achieves the smallest volume increase compared to previous works \cite{bloom2011convexhull, TanLG17}.
Bloom~\cite{bloom2011convexhull} runs appearance-preserving mesh simplification in primal space, but then needs to offset the face halfspaces to refit the data.
As a form of ablation of our technique, we consider replacing our primal-volume minimization during dual-hull simplification with the standard appearance-preserving quadrics \cite{GarlandH97}.
This often works surprisingly well (better than \cite{bloom2011convexhull,TanLG17} on this example), but still much larger than our output.

We treat convex hull simplification as a selection problem.
Selection is reasonable because we generally start with an abundance of choices and each choice is at least locally optimally oriented to the input data.
In Fig.~\ref{fig:Actaeon}, we compare again to \Vrep mesh simplification, which does not restrict output halfspaces to select from input and to $k$-DOPs which use prescribed (typically canonical) directions.

\begin{figure}
    \centering
    \includegraphics[width=\linewidth]{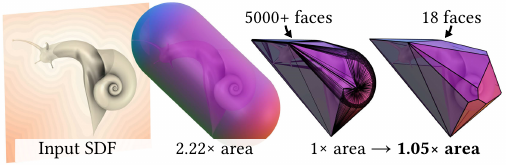}
    \caption{
    The zero-level set of (quasi) signed distance field from Inigo Quilez can be enclosed in a tighest-fitting but still loose capsule.
    Similar to \cite{Schott25} we extract points conservatively bounding the true convex hull of the implicit surface and simplify it.
    \label{fig:snail}
    }
    \includegraphics[width=\linewidth]{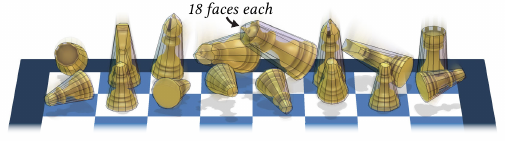}
    \caption{%
    Simplified convex hulls can be used for broadphase collision detection or directly as rigid body collision proxies.
    }
    \label{fig:chess}
    \includegraphics[width=\linewidth]{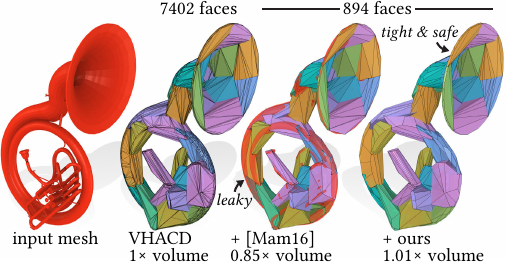}
    \caption{%
    Our simplification is a natural drop-in to approximate convex decomposition methods, e.g., VHACD \cite{Mamou2016}.
    Vertex-dropping simplification methods are common as a post-process for approximate convex decomposition methods, e.g., \cite{Mamou2016} and \cite{wei2022coacd}, but these cause each convex part to be \emph{smaller} at the risk of letting the contained shape leak through. Post processing with our simplification instead ensures tight and safe containment for the same plane budget.
    }
    \label{fig:sousaphone}
\end{figure}

\begin{figure*}
    \includegraphics[width=\linewidth]{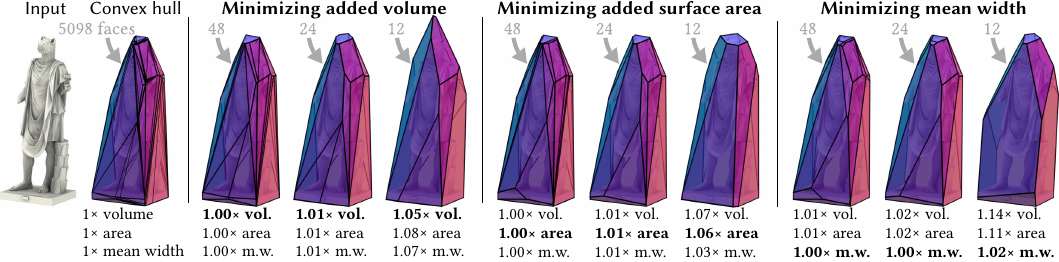}
    \caption{%
    Both are really tight, but we can choose to minimize volume, area, \newhl{or mean width}. Differences are a bit more pronounced at extreme simplifications, like this example down to 12 faces
    }
    \label{fig:hermanubis-broken}
    \includegraphics[width=\linewidth]{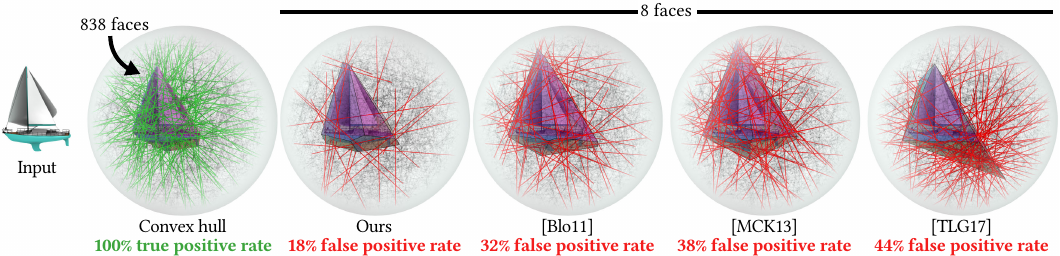}
    \caption{%
    \newhl{
    Minimizing added surface area is equivalent to minimizing false positive rate for random line intersections compared to the exact convex hull.
    We sample 1000 random lines in the tightest sphere containing the simplifications of various methods.
    Defining the true positive lines as those that  intersect the exact hull (green),
    false positive lines (red) intersect the simplification but not the exact hull.}
    }
    \label{fig:sailboat-lines}
\end{figure*}
Our technique is a form of greedy elimination. In Fig.~\ref{fig:motorcycle}, we compare to forward selection techniques.
For a very small number of target faces, greedy selection \cite{MuellerCK13} demonstrates bias to its initial three faces (suboptimally chosen).
Meanwhile, $k$-means clustering faces based on normals \cite{Schott25} often chooses faces poorly aligned with volume minimization.
Indeed, in Fig.~\ref{fig:k-means-killer} we pinpoint the issue with clustering normals with a painful pathological example.
Our volume minimizing simplification naturally adapts to the higher-resolution region by allotting more of its face budget there.
Meanwhile, $k$-means clustering of the input face normals \cite{Schott25} has the opposite of the desired effect: dense regions are undersampled and the budget is erroneously overspent on coarse regions, actually hurting the tightness in coarse, easy to approximate areas.

Similar to Schott et al.~\cite{Schott25}, we can build conservative and simplified convex hulls for implicit function inputs.
In Fig.~\ref{fig:snail}, we exploit the Lipschitz-ness of a quasi signed distance function to extract a conservative set of octree leaf cells.
The convex hull of the cell corners is guaranteed to contain the zero-level set.
Because the surface is smooth, its discrete approximation could be arbitrarily dense (5000+ faces shown here). We simplify to only 18 faces while only increasing the volume and surface area marginally compared to an oriented capsule (a cheap but loose bounding primitive).

Simplified convex hulls are useful for broadphase collision detection, such as determining if two rigidly deforming shapes might intersect.
Tight simplified hulls are also useful directly as simulation proxies, as seen in the dynamically colliding chess pieces in Fig.~\ref{fig:chess}.

For complex shapes, a single convex hull (exact or simplified) will often be too conservative and loose for most applications demanding more precise collision handling.
Fortunately, our simplification is a drop-in enhancement to existing approximate convex decomposition methods \cite{Mamou2016,wei2022coacd}.
In Fig.~\ref{fig:coacd}, we show that the default simplification used by \textsc{CoACD} \cite{wei2022coacd} is not safe, leaving gaps between decomposition cells.
A similar problem happens with the default simplification in \textsc{VHACD} \cite{Mamou2016} in Fig.~\ref{fig:sousaphone}, where part of the sousaphone sticks out of the simplified decomposition.
Post processing convex decomposition nodes with our simplification instead ensures tight and safe containment for the same plane budget and negligible change to the bounding volume.
For query acceleration (e.g., ray casting), area-minimizing bounding primitives are preferred \cite{PBRT3e}.
In Fig.~\ref{fig:hermanubis-broken}, we compare switching our cost function from volume added to surface area added and mean width increase.
Until extreme reductions, there is often not much difference. Minimizing surface area or mean width helps prevent long, thin spikes at extreme simplifications.

\newhl{
Minimizing volume is well motivated as minimizing the false positive rate for point containment queries, or as the leading term in small volumetric probe intersection tests (i.e., applications testing collisions with relatively small objects).
Meanwhile, minimizing surface area is often used as a heuristic for bounding volume hierarchy construction \cite{GoldsmithS87,MacDonaldB90} for accelerating ray intersections.
If a surface $\mathcal{S}$ lies in the unit sphere, and we sample $M$ random lines in the unit sphere, then:
\begin{align}
\text{area}(\mathcal{S})  = \lim_{M \rightarrow \infty} \frac{2 \pi K_\mathcal{S}}{M},
\end{align}
where $K_\mathcal{S}$ is the total number of line intersections with $\mathcal{S}$ \cite{LingMSJ25}.
For a convex shape $\mathcal{A}$ in the unit sphere, lines will generally intersect zero or two times, so we have:
\begin{align}
\text{area}(\mathcal{A})  = \lim_{M \rightarrow \infty} \frac{4 \pi N_\mathcal{A}}{M},
\end{align}
where $N_\mathcal{A}$ is the total number of lines intersecting $\mathcal{A}$.
If we have another convex shape $\mathcal{B} \supseteq \mathcal{A}$ (e.g., a simplified hull), then the conditional false positive rate (portion of intersections according to $\mathcal{B}$ which don't actually intersect $\mathcal{A}$) in the limit of $M$ lines is:
\begin{align}
\lim_{M \rightarrow \infty} \frac{N_\mathcal{B} - N_\mathcal{A}}{N_\mathcal{B}} = 1 - \frac{\text{area}(\mathcal{A})}{\text{area}(\mathcal{B})}.
\end{align}
When $\mathcal{A}$ is our input convex hull and $\mathcal{B}$ an enclosing simplification, this shows how minimizing the area of the simplification (or added area) reduces random line intersection false positives.
In Fig.~\ref{fig:sailboat-lines}, we show this visually in an example where we sample 1000 lines in a sphere containing various simplifications. Our method achieves the smallest added area, and unsurprisingly the smallest false positive rate.
}

\newhl{
If we switch the random lines to random planes, then the conditional false positive intersection rate is proportional to the mean width (a.k.a. mean caliper diameter or mean breadth):
\begin{align}
   \overline{w}(\mathcal{A}) =  
   \lim_{P \rightarrow \infty} \frac{2 Q_\mathcal{A}}{P}
\end{align}
where we consider $P$ random planes in the unit sphere and $Q_\mathcal{A}$ counts those that intersect $\mathcal{A}$. 
Width in a direction $\hat{\mathbf{u}} \in S^2$ is defined as the sum of the maximum distances to the normal plane at the origin in either direction (i.e., opposing support functions).
So, mean width is the average of all directions
\cite{moszynska2006selected}:
\begin{align}
\overline{w}(\mathcal{A}) &= \frac{1}{4 \pi} \int_{S^2} 
\left( h_\mathcal{A}(\hat{\mathbf{u}}) + 
h_\mathcal{A}(-\hat{\mathbf{u}}) \right)\, d\hat{\mathbf{u}},\\
h_\mathcal{A}(\hat{\mathbf{u}}) &= \max_{\mathbf{x} \in \mathcal{A} } \, \hat{\mathbf{u}} \cdot \mathbf{x}.
\end{align}
If $\mathcal{A}$ is a polyhedron then the mean width can be computed by summing over each edge $e$ its length $\ell_e$ times dihedral angle~$\theta_e$ \cite{Miles1969,achilleMSE}:
\begin{align}
\overline{w}(\mathcal{A}) = \frac{1}{4 \pi} \sum_{e}
\ell_e \theta_e.
\end{align}
As with volume change in Section~\ref{sec:geometry-change},
local increase in mean width can be computed efficiently (though, like surface area, not with rationals).
Minimizing mean width increase would likely reduce false positives when culling collisions with large objects (e.g., a ground plane).
}

\begin{figure}
    \centering
    \includegraphics[width=\linewidth]{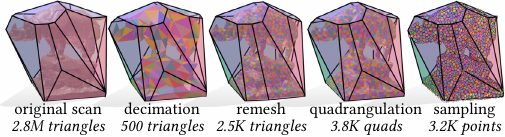}
    \caption{
    \label{fig:rhino}
    \newhl{Our simplification effectively chooses among the input's convex hull's faces. So, we inherit the stability of convex hulls. Various input perturbations result in rather stable final outputs.
    }
    }
\vspace{0.2cm}
    \centering
    \includegraphics[width=\linewidth]{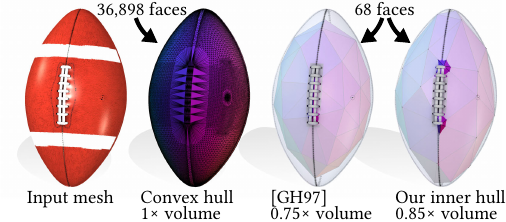}
    \caption{
    \label{fig:football}
    V-representations are good for inner approximations: we run our volume-minimizing simplification on this football's primal hull directly, producing a tighter inner approximation than standard mesh simplification \cite{GarlandH97}.
    }
\vspace{0.2cm}
    \centering
    \includegraphics[width=\linewidth]{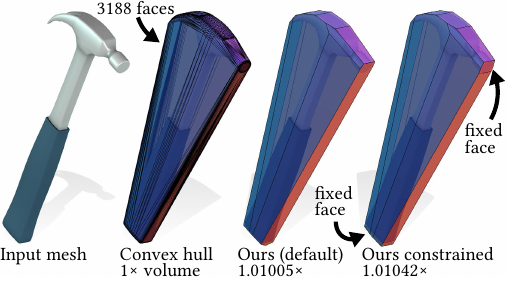}
    \caption{%
    \label{fig:hammer}%
    Our optimization-based simplification by default minimizes added volume (center-right), but can incorporate other losses and constraints, such as forcing specific faces in the output (right).
    }
\end{figure}

\newhl{%
In Fig.~\ref{fig:rhino}, we show how changing the input (and thus the input's convex hull) affects our final simplified hulls.
We run various geometry processing routines (decimation, remeshing, quadrangulation, blue noise sampling) on a dense triangle mesh scan. Each results in a different convex hull, and slightly different simplified output, exemplifying the stability of our approach.
}

For some collision proxy tasks, it might be useful to have an \emph{inner} approximation of an already roughly convex object (e.g., when avoiding false positives is more important than culling false negatives).
In Fig.~\ref{fig:football}, we run our volume-minimization on the primal hull's \Vrep directly. Compared to standard mesh-simplification \cite{GarlandH97}, ours is tighter to the original convex hull.

Finally, sometimes certain faces have special importance such as the base and striking-face of the hammer in Fig.~\ref{fig:hammer}. Our optimization can easily accommodate constrained faces to appear in the output by setting their removal as infinite cost.

\section{Limitations \& Conclusions}
\begin{figure}
    \centering
    \includegraphics[width=\linewidth]{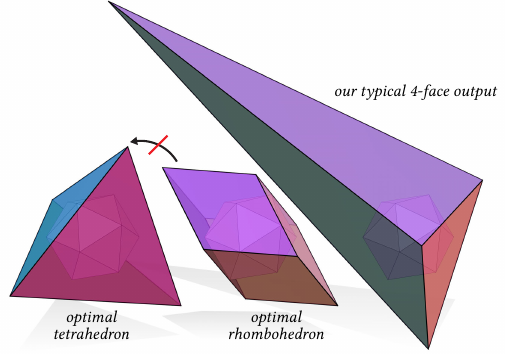}
    \caption{
    \label{fig:icosahedron}
    The globally optimal 4-faces simplification of an icosahedron is a regular tetrahedron (left).
    This is unreachable if our greedy path first finds the globally optimal rhombohedron (middle).
    Due to early indistinguishable symmetric breaking choices, we usually find a globally suboptimal tetrahedron for this extreme simplification example.
    }
\end{figure}

Finding the \emph{optimal} subset selection problem is NP-hard.
Like all previous works, our proposal is a heuristic. 
It enjoys well-studied theoretical bounds \cite{LopezR00,LopezR02,Reisner2001}
and performs well empirically.
Nevertheless, the hardness can sometimes exhibit itself in suboptimal solutions, especially noticeable for very symmetric shapes simplified to extremely coarse output.
In Fig.~\ref{fig:icosahedron}, we consider simplifying an icosahedron to four faces.
The optimal output would be a regular tetrahedron.
Due to symmetry in the priority queue costs, our algorithm must make early indistinguishable choices that preclude this output. Of 1000 trials of randomly rotated icosahedrons, the regular tetrahedron only appeared 0.6\% of the time.
The most frequent tetrahedron output (95.4\%) is irregular and larger.
Sometimes our greedy selection fails to reach four faces because it finds the globally optimal six-sided rhombohedron (1.2\%). The regular tetrahedron is unreachable from the rhombohedron because dropping any face makes it infinite volume (like a cube).

For future work,
it would be interesting to consider more elaborate elimination algorithms (e.g., beam search) or explore non-selection based solutions (perhaps for polishing).
Our implementation does not exploit parallelism: for future work, consider taking ideas from parallel mesh simplification (e.g., \cite{CignoniGGMPS05}).
We work with polyhedral hulls, but convex shapes can have smooth non-planar sides too.
Like the capsule in Fig.~\ref{fig:snail}, maybe exploiting smooth convexes like zonoids could provide tighter, cheaper-to-query bounding volumes.
Finally, we focused on 3D, but our foundational theory extends to higher dimensions, perhaps there are interesting applications for practical solutions there, too.

\subsection*{Acknowledgements}
``Actaeon'', ``Hermanubis-Broken'' and ``Rhinocéros'' from threedscans.
``Chess pieces on chessboard'' by HQ3DMOD, ``Owl'' by Claudio Naviglio, and ``Rustic wishing well'' by Laurie Annis, ``Claw hammer'' by Kollen Wasylean, ``Football'' by adobestock3d, ``Regatta Sailboat'' by Francesco Milanese from Adobe Stock.
``sousaphone'' by bmordue, ``Honda CB175 motorcycle'' by David Antalek on Sketchfab.
\censor{
My research is funded in part by NSERC Discovery (RGPIN–2022–04680), the Ontario Early Research Award program, the Canada Research Chairs Program, a Sloan Research Fellowship, the DSI Catalyst Grant program and gifts by Adobe Inc.
I am
grateful for the feedback of the Banh Mi weekly meeting attendees
at University of Toronto and delightful conversations with Ryan Schmidt and Qingnan Zhou. 
}

\bibliographystyle{eg-alpha}
\bibliography{references}

\appendix

\section{\newhl{Dual caps are generally non-convex}}
\label{sec:dual-caps-are-non-convex}
\begin{figure}[h!]
    \centering
    \includegraphics[width=\linewidth]{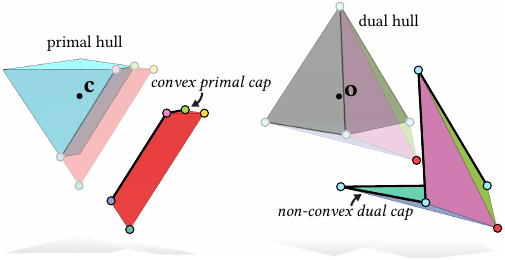}
    \caption{
    \newhl{
    Dropping a primal face gains the volume of the convex primal cap. The corresponding dual volume lost is non-convex.}
    \label{fig:dual-cap-non-convex}
    }
\end{figure}

\newhl{
Convex duals seem to maintain convexity for everything, so it's perhaps surprising that when we drop a primal halfplane, the ``primal cap'' volume \emph{gained} is convex while the ``dual cap'' volume \emph{lost} is generally non-convex.
The key reason for the non-convexity is that we haven't changed the center point about which the duality is defined, $\mathbf{c}$. 
We chose $\mathbf{c}$ to be deep inside the original primal hull; specifically it will not lie in the gained volume because that's entirely outside the original hull.
In Fig.~\ref{fig:dual-cap-non-convex}, consider dropping a face of the input primal hull (cyan), which gains the convex ``primal cap'' volume (red).
The dual about the primal center point $\mathbf{c}$ drops a ``dual cap'' volume made of faces incident on the respective dual point and the reverse of faces retriangulating that one-ring.
The dual cap is non-convex.}

\end{document}